\documentclass[prb,a4paper,reprint,showpacs,superscriptaddress,amssymb,floatfix]{revtex4-1}

\usepackage{bm}
\usepackage{graphicx}% \usepackage[notcite,notref]{tshowkeys}
\usepackage{latexsym}
\usepackage[dvips]{color} % this is required for text in color

\begin{document}

%%%%%%%%%%%%%%%% GENERAL DEFS %%%%%%%%%%%%
\def\Fbox#1{\vskip1ex\hbox to 8.5cm{\hfil\fboxsep0.3cm\fbox{%
  \parbox{8.0cm}{#1}}\hfil}\vskip1ex\noindent}  %%  {TEXT} in BOX

%%%%%%%%%%%%%%%%%%%%%%%  Refs to Eqs, Figs, Secs, Refs

\newcommand{\eq}[1]{(\ref{#1})}%%  requires \eq{label}
\newcommand{\Eq}[1]{Eq.~(\ref{#1})}%%  requires \eq{label}
\newcommand{\Eqs}[1]{Eqs.~(\ref{#1})}%%  requires \eq{label}
\newcommand{\Fig}[1]{Fig.~\ref{#1}}%%  requires \Fef{label}
\newcommand{\Figs}[1]{Figs.~\ref{#1}}%%  requires \Fef{label}
\newcommand{\Sec}[1]{Sec.~\ref{#1}}%%  requires \Fef{label}
\newcommand{\Secs}[1]{Secs.~\ref{#1}}%%  requires \Fef{label}
\newcommand{\Ref}[1]{Ref.~\cite{#1}}%%  requires \Fef{label}
\newcommand{\Refs}[1]{Refs.~\cite{#1}}%%  requires \Fef{label}

%%%%%%%%%%%%%%%%%%%%% Equation environment
\def\be{\begin{equation}}\def\ee{\end{equation}}
\def\bea{\begin{eqnarray}}\def\eea{\end{eqnarray}}
\def\bse{\begin{subequations}}\def\ese{\end{subequations}}
\newcommand{\BE}[1]{\begin{equation}\label{#1}}
\newcommand{\BEA}[1]{\begin{eqnarray}\label{#1}}
\newcommand{\BSE}[1]{\begin{subequations}\label{#1}}

\let \nn  \nonumber  \newcommand{\br}{\\ \nn}
\newcommand{\BR}[1]{\\ \label{#1}}
\def\hf{\frac{1}{2}}
\let \= \equiv \let\*\cdot \let\~\widetilde \let\^\widehat \let\-\overline
\let\p\partial \def\pp {\perp} \def\pl {\parallel}
\def\ort#1{\^{\bf{#1}}}
\def\Trans{^{\scriptscriptstyle{\rm T}}}
\def\x{\ort x} \def\y{\ort y} \def\z{\ort z}
\def\Re{\mbox{  Re}}
%%%%%%%%%%%%%%  Left-Right env:
\def\<{\left\langle}    \def\>{\right\rangle}
\def\({\left(}          \def\){\right)}
 \def \[ {\left [} \def \] {\right ]}

%%%%%%%%%%%%%%         Greeks
\renewcommand{\a}{\alpha}\renewcommand{\b}{\beta}\newcommand{\g}{\gamma}
\newcommand{\G} {\Gamma}\renewcommand{\d}{\delta}
\newcommand{\D}{\Delta}\newcommand{\e}{\epsilon}\newcommand{\ve}{\varepsilon}
\newcommand{\E}{\Epsilon}\renewcommand{\o}{\omega} \renewcommand{\O}{\Omega}
\renewcommand{\L}{\Lambda}\renewcommand{\l}{\lambda}
\renewcommand{\t}{\tau}
\def\r{\rho}\def\k{\kappa}
\def\t{\theta } \def\T{\Theta } \def\s{\sigma} \def\S{\Sigma}

%%%%%%%%%%%%%%%%%%% Bold, Calligraphic,  Gothic
%\newcommand{\B}[1]{{\bm{#1}}}%% Bold Roman & Greek Lower & Upper Case
\newcommand{\C}[1]{{\mathcal{#1}}}    %%   Calligrapfic Upper case
\newcommand{\BC}[1]{\bm{\mathcal{#1}}}%% Bold Calligrapfic Upper case
\newcommand{\F}[1]{{\mathfrak{#1}}}%% Fractur (Gothic) Lower & Uppers
%\newcommand{\BF}[1]{{\bm{\F {#1}}}}%    Bold Fractur (Gothic)

%%         Subscript and Superscript in Roman
\renewcommand{\sb}[1]{_{\text {#1}}}  %% sub-   for lower case
\renewcommand{\sp}[1]{^{\text {#1}}}  %% super- for lower case
\newcommand{\Sp}[1]{^{^{\text {#1}}}} %% Super- for Upper case
\def\Sb#1{_{\scriptscriptstyle\rm{#1}}}

\title{Propagation of thermal excitations in a cluster of vortices in superfluid $^3$He-B}

\author{J.J.~Hosio} \email{jaakko.hosio@aalto.fi}
\affiliation{Low Temperature Laboratory, P.O. Box 15100, FI-00076 AALTO, Finland}

\author{V.B.~Eltsov}
\affiliation{Low Temperature Laboratory, P.O. Box 15100, FI-00076 AALTO, Finland}

\author{R.~de~Graaf}
\affiliation{Low Temperature Laboratory, P.O. Box 15100, FI-00076 AALTO, Finland}

\author{M.~Krusius}
\affiliation{Low Temperature Laboratory, P.O. Box 15100, FI-00076 AALTO, Finland}

\author{J.~M\"akinen}
\affiliation{Low Temperature Laboratory, P.O. Box 15100, FI-00076 AALTO, Finland}

\author{D.~Schmoranzer}
\affiliation{Faculty of Mathematics and Physics, Charles University, Ke Karlovu 3, 121 16 Prague, Czech Republic}

\date{\today}

\begin{abstract}
We describe the first measurement on Andreev scattering of thermal excitations from a vortex configuration with
known density, spatial extent, and orientations in $^3$He-B superfluid. 
The heat flow from a blackbody radiator in equilibrium rotation at constant angular velocity 
is measured with two quartz tuning fork oscillators.
One oscillator creates a controllable density of excitations at $0.2~T_{\rm c}$ base temperature and the other records
the thermal response.
The results are compared to numerical calculations of ballistic
propagation of thermal quasiparticles through a cluster of rectilinear vortices.
We find good agreement which supports the current understanding of Andreev reflection.
%We describe the first measurement on Andreev scattering of thermal excitations from a vortex configuration with
%known density, spatial extent, and orientations in $^3$He-B superfluid. This configuration is created by
%rotating the $^3$He-B sample at constant angular velocity. We use two quartz tuning fork resonators
%embedded inside a blackbody radiator. One resonator creates a controllable density of excitations at $0.2~T_c$ base temperature and the other records
%the thermal response. The results are compared to numerical simulations of ballistic
%propagation of thermal quasiparticles through a cluster of rectilinear vortices.
%Our studies suggest that the current understanding of Andreev reflection is correct and it can be used as a quantitative tool to
%visualize vortices in the low temperature limit.

\end{abstract}


\maketitle %%

%Studies of turbulence in superfluid $^3$He and $^4$He have shown
%that quantum turbulence behaves in many ways similarly to classical turbulence. Therefore studying
%complex fluid motion in superfluids could help us to understand turbulence in classical fluids, which still has no comprehensive theory, as well.
%At very low temperatures, in the absence of normal fluid, turbulence in a superfluid condensate
%consists of a tangle of singly quantized topologically stable vortices with the same core size and circulation. Thus it is,
%at least on microscopic scales, simpler than turbulence in a normal fluid with eddies at different length scales.

%Studies of turbulence in superfluid $^3$He and $^4$He have shown
%that at large length scales quantum turbulence tends to mimic its classical counterpart. Therefore studying
%complex fluid motion in superfluids could help us to understand turbulence in classical fluids, which still lacks a comprehensive theory.
%At very low temperatures, in the absence of normal fluid, turbulence in a superfluid condensate
%consists of a tangle of singly quantized topologically stable vortices with the same core size and circulation. Thus it is,
%at least on microscopic scales, simpler than turbulence in a normal fluid with eddies at different length scales.
%Nevertheless, recent studies, both experimental and theoretical, have opened many challenging questions regarding
%quantum turbulence \cite{vinen}.
\section{INTRODUCTION}

Measurements of quantized vortices and attempts to develop methods for the visualization of different vortex configurations
has been central in superfluid studies. Recent interest has focused on quantum turbulence,
especially in the zero-temperature limit \cite{vinen}. In Bose-Einstein-condensed 
atom clouds motion of vortices has been recently investigated with optical means \cite{bagnato}. In the long-studied case of superfluid $^4$He, 
vortices can be detected, e.g., by ion trapping on vortex cores or second sound attenuation \cite{donnelly}, while
recent work on turbulence studies has been making use of transmission measurements of charged vortex rings \cite{Manchester} or 
trapping of micron-sized tracer particles in vortex tangles \cite{lathrop},
as well as by analyzing the drag force exerted on vibrating structures \cite{Skrbek}.
In superfluid $^3$He the traditional method to study vortices is nuclear magnetic resonance \cite{PLTP}. The superfluid flow
due to quantized vortices modifies the order parameter field and thus, the NMR signal. In uniform rotation a resolution
of a single vortex can be obtained \cite{ruutu,ab} in a measurement of the counterflow velocity at temperatures $T>0.5~T_c$.
At very low temperatures, in the limit $T/T_c\ll1$, the most powerful tool is the Andreev scattering of thermal excitations.
This technique has been developed and exploited at the University of Lancaster \cite{lanc_rev}.

Hitherto the Andreev scattering technique has only been used to detect turbulent vortex tangles, which for interpretation have been assumed to
be homogeneous and isotropic, but which in practice are of unknown density and
poorly known spatial extent. Thus, it has not been possible to compare theoretical predictions of heat transport in vortex systems directly to experimental results.
In this work we provide such a comparison and justify the use of the Andreev reflection technique as a visualization method of vortices in superfluid
$^3$He-B in the limit of vanishing normal fluid density.

\section{ANDREEV REFLECTION FROM VORTEX LINES}

In the ballistic regime of quasiparticle transport the mean free path of thermal excitations is longer than the dimensions of
the container. Therefore, thermal equilibrium is obtained via interaction of quasiparticles and container walls and
the collisions between excitations can be neglected.
In the presence of vortices, the superfluid flow field around the vortex lines can constrain the quasiparticle trajectories.

In the rest frame of the superfluid condensate the BCS dispersion relation $ E(\mathbf{p})$ is symmetrical and the minimum energy is the pressure dependent superfluid
energy gap $\Delta$. The standard picture of Andreev reflection considers an excitation moving towards an
increasing energy gap \cite{andreev}. In $^3$He-B the superfluid flow field modulates the minimum in the excitation spectrum.
Using the notation by Barenghi \textit{et al.} \cite{barenghi1}, the energy $E$ of the excitation with momentum $\mathbf{p}$ in the flow field around a vortex is given by
\begin{equation}
E(\mathbf{p})=\sqrt{\mathbf{\e}_p^2+\Delta^2} + \mathbf{p}\cdot\mathbf{v_s},
\end{equation}
where $\mathbf{\e}_p=p^2/2m^*-\e_F$ is the effective kinetic energy of the excitation measured with
respect to the Fermi energy $\e_F$ and $p=|\mathbf{p}|$. Excitations with $\mathbf{\e}_p>0$ are called quasiparticles and excitations with
$\mathbf{\e}_p<0$ are called quasiholes. For quasiparticles the group velocity $\mathbf{v}_{\rm g}(E)=dE/d\mathbf{p}$ is parallel to 
the momentum $\mathbf{p}$ whereas for
quasiholes it is antiparallel. Our experiments are performed at the 29~bar pressure, at which the effective
mass $m^* \approx5.42~m_3$, where $m_3$ is the mass of a $^3$He atom. The superfluid velocity $\mathbf{v}_{\rm s}$ is proportional to the gradient of the phase
$\varphi$ of the order parameter, i.e., $\mathbf{v}_{\rm s}=\hbar/(2m_3)\nabla\varphi$. If we consider a vortex oriented along the $z$ axis in cylindrical
coordinates ($r$,$\phi$,$z$) this becomes $\mathbf{v}_{\rm s}=\kappa/(2\pi r)\hat{\phi}$, where $\kappa=h/2m_3\approx6.62\times10^{-4}$ cm$^2$/s is the circulation 
quantum and $\hat{\phi}$ the azimuthal unit vector.

The consequence of the interaction term $\mathbf{p} \cdot \mathbf{v}_{\rm s}$ is that an excitation traveling 
at constant energy may not find a forward-propagating state due to the superflow gradient $\nabla \mathbf{v}_{\rm s}$ along
the flight path. When the excitation reaches the minimum of the spectrum the group velocity changes sign and it
retraces its trajectory as an excitation on the other side of the minimum. In other words, a quasiparticle Andreev reflects 
as a quasihole and vice versa with a very small momentum transfer \cite{lanc93}.

Let us consider a beam of excitations incident on a single straight vortex. On one side of the vortex the flow parallel to the group velocity of the excitation
reflects quasiparticles and on the other side the antiparallel flow reflects quasiholes. An excitation is Andreev reflected if its energy satisfies
$E \leq \Delta+p \kappa/(2\pi b)\sin \theta$. Here $\theta$ is the inclination angle  of the excitation trajectory measured with respect to the 
vortex line and $b$ the impact parameter. At temperature $T$ the mean excitation energy is $\tilde{E}=\Delta+k_B T$. In our experiments $k_B T\sim0.1~\Delta$ 
and the momentum $p$ is close to the Fermi momentum
$p_F\approx9.26\cdot10^{-25}$ ~kgm/s, so a typical excitation is reflected if $b<5 p_{F} \kappa \sin \theta /(\pi \Delta)$. 
For an excitation with $\theta\approx\pi/4$ this translates to $\sim1\mu m$, which is about two orders of magnitude larger than the coherence length 
$\xi\approx15~$nm and the vortex core radius. Thus, in a typical experimental situation the probability 
of scattering off a vortex core is negligible compared to the cross section for Andreev scattering from the flow field around the vortex.
%In our measurements the typical inter-vortex distance $l>0.1$~mm is

\section{EXPERIMENTAL METHODS}

In our experiment we study the heat transported by excitations through a cluster of vortices. Bradley and coworkers did a similar
measurement with a vortex tangle as the structure reflecting excitations \cite{bradley}. Our experiment is made in a fused quartz cylinder filled
with {$^3$He-B}. The cylinder is divided in two parts with a 0.7~mm thick quartz division plate. The lower part consists of a 30~mm long, 3.5~mm inner diameter tube, which
opens to a heat exchanger made out of sintered silver. The sinter provides good thermal contact with the nuclear cooling stage so that
the superfluid $^3$He below the 0.3 mm orifice can be cooled down to below 0.15~$T_\textrm{c}$. The upper part can be modeled as a blackbody radiator (BBR), an enclosure
with a weak thermal link to the outside superfluid $^3$He via a small orifice in the division plate \cite{lanc84}. Our BBR consists
of a 12~cm long section of the quartz tube with 6~mm inner diameter. This volume is furnished with
two mechanical resonators, one acting as a thermometer and the other as a heater. The heater is used for generating a beam of ballistic quasiparticles
through the orifice.

Our resonators are commercial quartz tuning forks, which have recently been characterized for probing quantum fluids \cite{our_fork,lanc_fork}. The forks
are made of piezoelectric quartz crystals with electrodes deposited on the surface. When driven with alternating voltage, the
two prongs of the fork oscillate in anti-phase producing a current $I$, which is proportional to the prong velocity $v_{\rm p}$.
The heater fork signal is amplified with a room-temperature I/V converter \cite{skyba_conv} before being fed to a two phase lock-in amplifier.
This was found to be important to reduce capacitive losses in the signal line, and thus to measure accurately the signal amplitude which is proportional
to the power generated by the fork.
The thermometer fork has 32~kHz resonance frequency, a prong cross section of 0.10~mm $\times$ 0.24~mm and a length 2.4~mm. The heater
fork has a higher resonance frequency, 40~kHz, to prevent any interference between the forks. The prongs of the heater are 2.9~mm long and the
cross section is 0.36~mm$\times$0.44~mm.

%%%%%%%%%%%%%%%%%%%%%%%%%%%%%%%%%%%%%%%%%%%%%%%%%%%%%%%%%%
\begin{figure}[t!]
\begin{center}
\centerline{\includegraphics[width=0.95\linewidth]{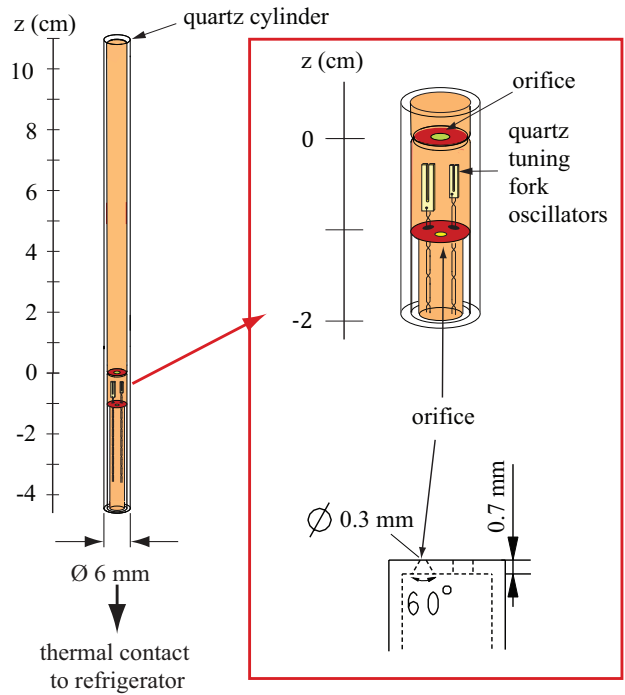}}
%\vspace{-7mm}
\caption{The experimental setup. The upper experimental volume modeled as a black body radiator
is separated from the heat exchanger volume at the bottom by a division plate with a conical orifice with 0.3~mm diameter. 
The upper division plate with a 0.75~mm diameter aperture is not relevant to the measurement described here.
The BBR houses two quartz tuning fork oscillators one acting as heater, the other as thermometer.  }
\label{setup}
\end{center}
\vspace{-7mm}
\end{figure}
%%%%%%%%%%%%%%%%%%%%%%%%%%%%%%%%%%%%%%%%%%%%%%%%%%%%%%%%

In our temperature range the resonance width of the tuning fork depends only on the damping from ballistic quasiparticles. The dependence
of the linewidth $\Delta f$ on temperature and prong velocity is given by
\begin{equation}
\Delta f=\Delta f_{\rm{int}}+a e^{-\Delta/k_BT}(1-\lambda \frac{p_F}{k_B T}v_{\rm p}) ,
\label{width}
\end{equation}
where $\lambda$ is a geometrical factor close to unity \cite{skyba}. The second term in the parenthesis is due to Andreev reflection of thermal
quasiparticles from the potential flow field created by the fork prongs moving the liquid around them. In our experiments $v_p$ is small
and the velocity-dependent term in Eq. (\ref{width}) can be neglected. Thus,
calibrating the fork to act as a thermometer requires determining only the geometry-dependent factor $a$.
The thermometer is calibrated at 0.33~$T_c$ against a $^3$He-melting curve thermometer, which is thermally coupled to the
heat exchanger. Our calibration gives $a\approx$17500~Hz for the detector. The intrinsic
damping of the fork was measured to be $\Delta f_{int}\approx14$~mHz at $T\sim$~10~mK in vacuum, which translates into a quality factor $Q\sim2\cdot10^6$.

The rough surface of the sinter with a grain size close to the vortex core diameter provides excellent
spots for vortices to nucleate. Thus, the critical rotation velocity  $\Omega_\textrm{c}$ for vortex formation is lower than 0.1~rad/s in
the bottom section of the long quartz tube.
In the equilibrium vortex state in uniform rotation the sample becomes filled with rectilinear vortices oriented along the rotation axis. 
To create the vortex array, which Andreev reflects a part of the heat back to the BBR, we rotate our system at constant velocity around the axis
of the container tube.
The vortex density in the 
equilibrium state is determined by
minimization of the free energy in the rotating frame and is given by the solid-body-rotation value $n_v=2\Omega /\kappa$.
The array is isolated
from the container wall by a narrow annular vortex-free layer. The width of the vortex-free region $\sqrt{\kappa/(\sqrt{3}\Omega)}$ 
is only slightly larger than the intervortex distance \cite{krusius}.

%%%%%%%%%%%%%%%%%%%%%%%%%%%%%%%%%%%%%%%%%%%%%%%%%%%%%%%%%%
\begin{figure}[schem]
\begin{center}
\centerline{\includegraphics[width=0.95\linewidth]{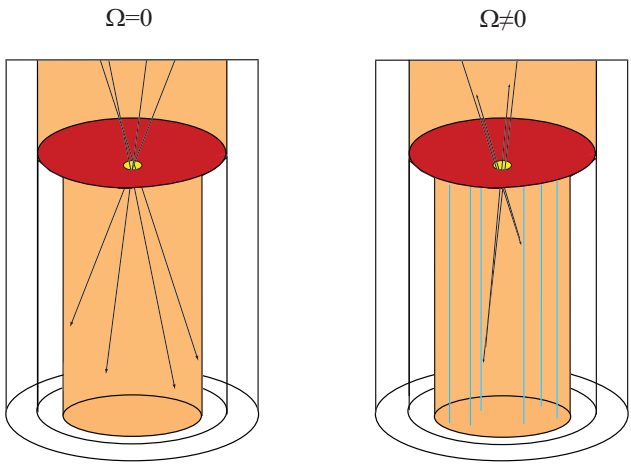}}
%\vspace{-7mm}
\caption{Sketch of the experiment. In the system at rest ($\Omega$=0) all the excitations which do not migrate back to the blackbody radiator
due to diffuse scattering from the walls are thermalized in the
heat exchanger at the bottom. In rotation ($\Omega\neq0$), part of the beam is Andreev reflected from the cluster of vortices.}
\label{schem}
\end{center}
\vspace{-7mm}
\end{figure}
%%%%%%%%%%%%%%%%%%%%%%%%%%%%%%%%%%%%%%%%%%%%%%%%%%%%%%%%

To make sure that we have the equilibrium number of vortices in the container we first rapidly
increase the rotation velocity to some value which is higher than the target velocity for the measurement. Then we go to the final velocity and wait for the
system to settle to the equilibrium vortex state after the annihilation of the extra vortices and the slow relaxation of the vortex array. 
In our experimental conditions the relaxation takes about one hour.

All the power entering our experimental volume modeled as a blackbody radiator must leave through the hole at the bottom as a flux of energy-carrying
excitations. Assuming thermal equilibrium inside the BBR the power is given by
\begin{equation}
\label{p1}
\dot{Q}(\Omega)=\int N(E) v_g(E) Ef(E)\C{T}d E d x d y d \phi d \theta,
\end{equation}
where $N(E)$ and $f(E)$ are the quasiparticle density of states and the Fermi distribution function, respectively. In the limit $k_B T\ll\Delta$ the latter
reduces to the Boltzmann distribution $f(E)=e^{-E/k_BT}$. The transmission function $\C{T}=\C{T}(E,x,y,\phi,\theta,\Omega)$ is equal to one
if an excitation leaving the BBR (at position $(x,y)$ on top of the orifice to direction $(\phi,\theta)$) reaches the sinter and zero if
it is reflected back. The integration goes over the cross section of the orifice, $\phi \in (0,2\pi)$, $\theta \in (0,\pi/2)$  and $E \in (\Delta,\infty)$.
The power generated inside the radiator can now be expressed as the sum of the $\Omega$-dependent residual heat leak $\dot{Q}_{hl}$ to the BBR and the direct
power $P_{gen}$ from the excitations produced by the heater fork
\begin{equation}
\label{p2}
\dot{Q}_{hl}(\Omega)+P_{gen}=\frac{4\pi k_B p_F^2}{h^3}Te^{-\frac{\Delta}{k_BT}}(\Delta+k_B T) A_h(\Omega).
\end{equation}
Here $A_h(\Omega)$ is the effective area of the orifice, which gets smaller when part of the excitations is scattered back to the BBR. We omit the flow
of excitations from the thermal excitation bath in the heat exchanger volume since the quasiparticle density there is at
least three orders of magnitude lower than inside the BBR, which is at $0.20~T_\textrm{c}$.

%%%%%%%%%%%%%%%%%%%%%%%%%%%%%%%%%%%%%%%%%%%%%%%%%%%%%%%%%%%%%%%%%%%%%%%%%%%%%%%%%%%%%%%%%%%%%%%%%%%%%%%%%%%%%%%%%%%%%%%%%%%%%%%%%%%%%%%%

\section{STEADY STATE MEASUREMENTS}

In the measurement the heater fork is driven to create the desired excitation beam corresponding to power $P_{gen}$ leaving the radiator.
By controlling the rotation velocity, and thus the vortex density, we can control the fraction
of Andreev reflected excitations. As illustrated in Fig. \ref{schem}, the flow field created by the vortices reflects part of the beam
back to the radiator by Andreev scattering. As a consequence, the temperature above the orifice increases
more than with the same applied heating in the absence of vortices.
The fraction $\nu$ of heat reflected back into the radiator, which we call the
reflection coefficient, can be obtained from Eq. (\ref{p2}) as
\begin{equation}
\label{rc}
\nu(\Omega)=1-\frac{A_h(\Omega)}{A_h(0)}.
\label{rcoeff}
\end{equation}
There is also a vortex cluster inside the BBR, which may cause
a small temperature gradient along the cylinder. Nevertheless, the main thermal resistance is across the orifice and
Eq. (\ref{rc}) is valid as long as there is thermal equilibrium inside the radiator.

%%%%%%%%%%%%%%%%%%%%%%%%%%%%%%%%%%%%%%%%%%%%%%%%%%%%%%%%%%
\begin{figure}
\begin{center}
\centerline{\includegraphics[width=0.9\linewidth]{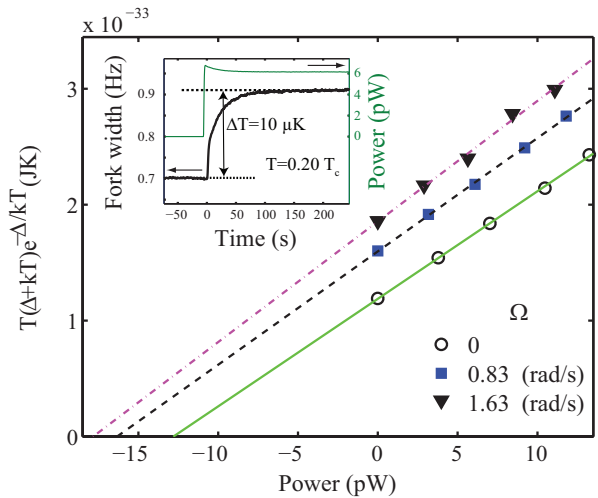}}
%\vspace{-7mm}
\caption{Temperature dependent part of the power leaving the blackbody radiator as a function of heating power at three different rotation
velocities. The temperature is obtained from the linewidth of the detector fork. The shown data points are averaged from 
data measured for about 10 minutes at each power. The intercept of the linear fit with the power axis gives the residual heat leak to the sample and the
effective area is given by the inverse of the slope. The slope, the heat leak and the scatter in the data all increase with increasing
angular velocity. The inset shows an example of a detector response to a heating pulse starting at time $t=0$. }
\label{meas}
\end{center}
\vspace{-7mm}
\end{figure}
%%%%%%%%%%%%%%%%%%%%%%%%%%%%%%%%%%%%%%%%%%%%%%%%%%%%%%%%

At each rotation velocity, we apply different power inputs to the radiator and measure the corresponding equilibrium temperature
with the thermometer fork. By plotting all the temperature-dependent parts in Eq. (\ref{p2}) as a function of power $P_{gen}$ we get
a straight line (Fig. \ref{meas}). From the inverse slope of the line we get
the effective area $A_h$ and from the intercept with the power axis the heat leak $\dot{Q}_{hl}$.

The measurement with no vortices
gives $A_h(0)\approx0.020$~mm$^2$. This is less than a half of the geometrical area $A_g$ of the orifice. The
main reason is the diffusive backscattering of the excitations from the walls of the 0.7~mm thick division plate and the quartz tube below it.
In any case, the absolute value of the effective area is not an important issue since we are only
interested in the relative change in Eq. (\ref{rc}).
The heat leak $\dot{Q}_{hl}$ varies from 12~pW at $\Omega=0$ to 18~pW at $\Omega=$1.8~rad/s. At high rotation velocities
the rotation-induced heat leak fluctuates with variations of about 1~pW. The rotation velocities used in
the measurements had to be carefully selected, since mechanical resonances at certain velocities cause increased and temporally varying heating. 
Figure \ref{res} shows the reflection coefficient as a function of the rotation velocity.
In the measured rotation velocity range the dependence of  $\nu$  on the vortex density is approximately linear. 
We believe that the main source of scatter in the experimental data comes from the variation
of the power calibration of the heater fork.

In the measurements the rotation velocity fluctuates on the level of $\Delta \Omega \leq$ 0.01~rad/s. Therefore, it is possible that we
create helical perturbations on vortex lines, which can end up increasing the total vortex length and decreasing the polarization in our vortex cluster. By modulating
the rotation velocity at different frequencies and amplitudes we can study whether the presence of these perturbations, which are called Kelvin waves,
affects the reflection coefficient. We find that even an order of magnitude larger modulation amplitude compared to the highest noise peaks in the rotation
velocity barely affects the fraction of transmitted heat flux. Thus, we believe it is safe to omit the effect of Kelvin waves in our analysis.
Our preliminary measurements at very large modulation amplitudes, however, show a decreasing fraction of transmitted heat, as expected if Kelvin waves
are generated. In the future, we are hoping to utilize these techniques to study Kelvin waves in more detail.

\section{TRANSIENT MEASUREMENTS}
To test whether our blackbody radiator works as expected, we can analyze how the system reaches thermal equilibrium when
the heater is suddenly switched on. The expected time constant for the thermal relaxation is $\tau=RC$, where
the thermal resistance across the orifice $R=(d\dot{Q}/dT)^{-1}\propto A_h^{-1}$ and the heat capacity $C$ is given approximately by \cite{bauerle}
\begin{equation}
C=k_B \sqrt{2\pi}N_F\left(\frac{\Delta}{k_BT}\right)^{\frac{3}{2}} e^{-\frac{\Delta}{k_BT}}\left(\Delta+\frac{21}{16}k_B T\right) V.
\end{equation}
Here $V\approx3.4$~cm$^3$ is the volume of the BBR and $N_F$ the density of states at the Fermi level.
The measured time constant is about 25~s (see inset in Fig. \ref{meas}), which is in a good
agreement with the expected time constant ($\sim32$~s) obtained using the effective area from the calibration described above. This analysis also proves that
practically all the heat capacity of the system is in the bulk superfluid $^3$He. The possible error sources in the effective area are the
small statistical error in the determination of the slope (see Fig. \ref{meas}) and uncertainties in the power calibration, temperature calibration and the
value of the gap \cite{todo}. The reflection coefficient $\nu$ has a weak logarithmic dependence on the parameter $a$ and depends on $\Delta$ only through 
the temperature calibration [Eq. (\ref{width})].
The power calibration, if time-independent, has no effect on $\nu$. We estimate the overall systematic uncertainty of the measurements
shown in Fig. \ref{res} to be of the order of the scatter in the data.

%%%%%%%%%%%%%%%%%%%%%%%%%%%%%%%%%%%%%%%%%%%%%%%%%%%%%%%%%%
\begin{figure}
\begin{center}
\centerline{\includegraphics[width=0.9\linewidth]{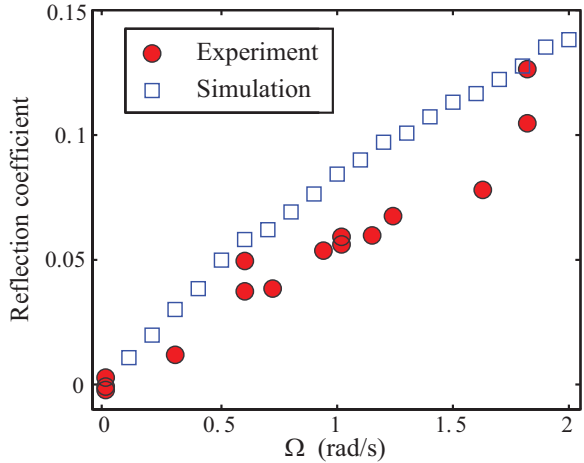}}
%\vspace{-7mm}
\caption{The fraction $\nu$ of the heat Andreev reflected back into the blackbody radiator. The temperature inside the radiator is $0.20~T_\textrm{c}$.
The simulation points are obtained by integrating Eq. (\ref{p1}) numerically and solving equations (\ref{p2}) and (\ref{rc}) for $\nu$.}
\label{res}
\end{center}
\vspace{-7mm}
\end{figure}
%%%%%%%%%%%%%%%%%%%%%%%%%%%%%%%%%%%%%%%%%%%%%%%%%%%%%%%%

\section{SIMULATION CALCULATIONS}

In our numerical simulations we calculate the transmission function $\C{T}$ for our geometry at different rotation velocities and solve the integral
in Eq. (\ref{p1}) numerically using Monte Carlo integration with importance sampling. 
The simulations use the exact geometry of our experimental setup including the thickness and the shape of the radiator orifice. Instead of solving for
the full equations of motions for excitations, which would require too much computing power, only the vortices for which the impact parameter
of the excitation is small enough to allow Andreev reflection are considered. We do not assume perfect retro-reflection but take into account the small Andreev
reflection angle \cite{barenghi1} $\Delta \varphi= \hbar p_F^{-1} \sqrt{\pi/(3\xi b)}$. 
Figure \ref{res} shows the reflection coefficient as a function of the rotation velocity from the numerical simulations. The result is in a good
agreement with the measurements.

In the simulations diffuse scattering from the container walls is assumed based on
the results in Ref. \onlinecite{scat}. In reality, part of the excitations experience
Andreev reflection at the container surfaces \cite{kieselmann}. Including this effect would lead to a decrease of the reflection coefficient
in the numerical simulations, as confirmed by our calculations with an unrealistic steplike order-parameter variation,
and improve the agreement between the simulations and the measurements. The full
accounting for the detailed reflection processes at diffusely scattering surfaces, including a slight particle-hole anisotropy \cite{zhang}, would 
complicate the transport calculation and we neglect these effects, arguing as follows. Our simulations with fully diffusive scattering yield a ratio $A_h(0)/A_g$ which
agrees well with the known $A_g$ and $A_h(0)$ obtained from our measurement. This is neither the case for the model with a steplike order parameter variation at surfaces
nor for specular scattering, since to produce the measured $A_h(0)$ the former would require $A_g$ to be roughly two times larger and the latter less than a half of the known 
geometrical area.
Moreover, quasiclassical calculations with
a smooth model order-parameter suppression profile \cite{zhang,zhang2} show that the real  Andreev reflection probability at a diffusely
scattering surface is only a small fraction of that in the step model. Finally, even though the details of the wall scattering processes do have an effect on $A(\Omega)$, they
do not affect the reflection coefficient $\nu$ significantly compared to other uncertainties in our model.

Recent numerical studies \cite{barenghi2,barenghi3} indicate that especially for dense vortex structures, the total reflecting
"Andreev shadow" is not necessarily the sum of shadows of single vortices. Our clusters are relatively sparse and moreover, after the first diffuse
scattering from the walls the probability for the excitation to migrate back to the radiator is not sensitive to small changes in its trajectory.
Thus, we believe that our somewhat simplified model reproduces the real experimental situation with good accuracy. The simulations
were tested at different hole radii and positions of the hole on the division plate. We find that the largest effect on the final result comes from the uncertainty of the radius: 
increasing or decreasing it by 50\% changes the reflection coefficient by approximately $\pm10\%$.

%\emph{At higher modulation amplitudes, however, a clear increase of the reflection coefficient can be seen. Our earlier experiments \cite{spindown} showed
%that the vortex multiplication and annihilation happens at much longer timescales than one period in our modulation measurements. Thus, the number of
%vortices should remain constant and the increase of $\nu$ must reflect increase in the total vortex length due to Kelvin waves. Nevertheless, more experimental and
%numerical studies are needed to confirm this result.}
\section{CONCLUSIONS}
In conclusion, we describe the first measurement of the interaction between thermal excitations and quantized vortices in a well-defined configuration. Numerical simulations
reproduce the experimental results within the margin of uncertainty in the model.
Quasiparticle beam techniques are currently the most popular measuring method of vortices in $^3$He-B below $0.2~T_\textrm{c}$.
Our work provides a rigorous quantitative basis for their use and further development for direct visualization purposes.

%%%%%%%%%%%%%%%%%%%%%%%%%%%%%%%%%%%%%%%%%%%%%%%%%%%%%%%%

\section*{ACKNOWLEDGEMENTS}
This work is supported by the Academy of Finland (Centers of Excellence Programme  2006-2011 and grant 218211) and the EU 7th Framework Programme (FP7/2007-2013,  grant 228464 Microkelvin).
JH acknowledges financial support from the V\"{a}is\"{a}l\"{a} Foundation of the Finnish Academy of Science and Letters and
useful discussions with R. Haley, N. Kopnin, Y. Sergeev, and P. Skyba.

%%%%%%%%%%%%%%%%%%%%%%%%%%%%%%%%%%%%%%%%%%%%%%%%%%%%%%%%

\end{document}